\documentclass[a4paper, 11pt]{article}
\usepackage{comment} % enables the use of multi-line comments (\ifx \fi) 
\usepackage[nointegrals]{wasysym}
\usepackage{fullpage} % changes the margin
\usepackage{graphicx}
\usepackage{cite}
\usepackage{bm}
\usepackage{float}
\usepackage{amsmath}
\usepackage[utf8]{inputenc}
\usepackage{amssymb}
\usepackage{mathrsfs}
\usepackage{wrapfig}
\usepackage{enumitem}
\usepackage[english]{babel}
\usepackage{comment}
\usepackage{bbold}  % identity with \mathbb{1}
\usepackage{multicol}
\usepackage[usenames,dvipsnames]{xcolor} 
\usepackage{xcolor,hyperref}
\usepackage{setspace}
\usepackage{abstract}
\usepackage{sectsty}
\sectionfont{\large}
\usepackage{tablefootnote}

\newcommand\myfontsize{\fontsize{10pt}{12pt}\selectfont}
\usepackage[T1]{fontenc}
\usepackage[margin=0.94in]{geometry}
\begin{document}
\null\hfill MPP-2023-134\; \\
\null\hfill LMU-ASC 21/23\; \\
\null\hfill CERN-TH-2023-116
\vspace*{\fill}
\begin{center}
    \LARGE\textbf{\textsc{ \textcolor{Black}{On the Ghost Problem of Conformal Gravity} }}

    \normalsize\textsc{Anamaria Hell$^{\eighthnote}$, Dieter Lüst$^{\eighthnote,\;\twonotes}$,  George Zoupanos$^{\twonotes,\;   \flat\quarternote,\; \sharp\halfnote,\;  \fullnote }$}
\end{center}

\begin{center}
    $^{\eighthnote}$\textit{Arnold Sommerfeld Center for Theoretical Physics,\\
Ludwig–Maximilians–Universität München\\
Theresienstraße 37, 80333 Munich, Germany}\\
$^{\twonotes}$\textit{Max–Planck–Institut für Physik (Werner–Heisenberg–Institut)\\
Föhringer Ring 6, 80805 Munich, Germany}\\ 
$^{\flat\quarternote}$\textit{Physics Department, National Technical University,\\ Zografou campus
157 80 Athens, Greece}\\
$^{\sharp\halfnote}$\textit{Theory Department, CERN,\\ Geneva 1211, Switzerland
}\\
$^{\fullnote}$\textit{Institut für Theoretische Physik der Universität Heidelberg,\\Philosophenweg 16, 69120 Heidelberg, Germany}
\end{center}
\thispagestyle{empty} 

\renewcommand{\abstractname}{\textsc{\textcolor{Black}{Abstract}}}

\begin{abstract}
 We study the metric perturbations around the de Sitter and Minkowski backgrounds in Conformal Gravity. We confirm the presence of ghosts in both cases. In the de Sitter case, by applying the Maldacena boundary conditions -- the Neumann boundary condition and the positive-frequency mode condition -- to the metric, we show that one cannot recover a general solution for the perturbations. In turn, alongside the Neumann boundary condition, we derive an additional condition with which the perturbations of conformal gravity and dS perturbations of Einstein gravity with cosmological constant coincide. We further show that the Neumann boundary condition does not lead to a general solution in Minkowski space. Conversely, we derive the alternative boundary conditions, with which we attain an agreement between the perturbations of conformal and Einstein gravity in full generality, thus removing the ghost of conformal gravity.

\end{abstract}
\begin{comment}
\vfill
\small
\href{mailto:Hell.Anamaria@physik.uni-muenchen.de}{\textsc{Hell.Anamaria@physik.uni-muenchen.de}}\\
\href{mailto:dieter.luest@lmu.de}{\textsc{dieter.luest@lmu.de}}\\
\href{mailto:george.zoupanos@cern.ch}{\textsc{george.zoupanos@cern.ch}}
\end{comment}
\vspace*{\fill}
\clearpage
\pagenumbering{arabic} 
\newpage

\section{  {\textcolor{Black}{\Large \textbf{\textsc{Introduction}}}}}

To date, Einstein's gravity has reached numerous successes, ranging from perturbative confirmations such as the precession of the perihelia of Mercury and deflection of the starlight to non-perturbative ones, such as the Big Bang Nucleosynthesis. Yet, from the field theory perspective, it is incomplete. Although at one loop it is shown to be renormalizable \cite{tHooft:1974toh}, this property is lost as soon as one considers the higher order corrections \cite{Goroff:1985th}. 

Higher-derivative theories of gravity \cite{Stelle:1976gc, Stelle:1977ry, Boulware:1983td, David:1984uv, Deser:2007vs, Buchbinder:1987vp, tHooft:2011aa, Alvarez-Gaume:2015rwa, Horowitz:1984wv, Park:2012ds, Lu:2011ks}, on the other hand, are advocated to be power-counting renormalizable\footnote{However, see also \cite{Capper:1975ig}, where an argument against renormalizability was given.\hfill } and asymptotically free \cite{Stelle:1976gc, Fradkin:1981iu, Julve:1978xn}.  As such, these theories could play a role in the UV completion of gravity \cite{Adler:1982ri, tHooft:2011aa}. 

\textsc{Conformal Gravity} (CG)   \cite{Weyl:1918ib, Weyl:1919fi, Bach} --  model whose action is made out of the square of the Weyl tensor -- takes a predominant role among these theories. Its action is characterized by the invariance under the conformal transformations -- local (\textit{or space-time dependent}) re-scalings of the metric -- making it dependent only on angles and not distances. 

Over the years, the interest in this theory has been vast. Formulated as a gauge theory, it has been an important ingredient in the construction of supergravity theories  \cite{Kaku:1977pa, Fradkin:1985am,  Kaku:1978ea, Kaku:1978nz, deWit:1980lyi, Bergshoeff:1980is, Liu:1998bu,  Andrianopoli:2014aqa, DAuria:2021dth, Ferrara:2018wqd, Ferrara:2020zef}, and non-commutative geometry \cite{Chamseddine:1996zu, Manolakos:2021rcl, Manolakos:2019fle}. It also emerges from the twistor string theory \cite{Berkovits:2004jj}, and appears in different renormalization prescriptions of the gravitational theories, such as holographic, Kounterterm, and conformal renormalization \cite{ Sen:2012fc, Grumiller:2013mxa, Miskovic:2009bm, Anastasiou:2022wjq, Anastasiou:2020zwc }. In addition, CG has also been applied in Cosmology and Black-Hole (BH) physics.  It has been advocated to explain the flat galaxy rotation curves without the need for dark matter \cite{Mannheim:2012qw} (although recently there have been tensions with these claims -- see eg. \cite{Hobson:2021vzg}), and may also provide for astrophysical BH candidates \cite{Meng:2022kjs, Momennia:2019edt}.
 
Regardless of its intriguing properties and applications, CG suffers from a pathology. It describes two vector \textit{degrees of freedom (dof)}, and four tensor ones \cite{Riegert:1984hf}. However, due to the presence of higher derivatives, two tensor modes among these are the Ostrogradsky ghosts \cite{Ostrogradsky:1850fid} -- they
cause the energy to be unbounded from below \cite{Boulware:1983td}. Due to these modes, classically, the system exhibits a linear instability, while upon quantization, they lead to the violation of unitarity.

Notably, in \cite{Maldacena:2011mk}, Maldacena has conjectured that the key to resolving this problem lies in the boundary conditions, with which CG would reduce to \textit{Einstein gravity with a cosmological constant} ($E\Lambda$) in Euclidean AdS and dS space (see also \cite{Anastasiou:2016jix, Anastasiou:2020mik}). By introducing the Neumann Boundary Condition, in which the first derivative of the metric vanishes at the boundary, together with the Positive Frequency Condition -- the requirement that at very early times only the positive frequency modes are present -- the four tensor modes would recombine in such way that only two would remain, which would agree with the two tensor modes of the de Sitter space. While such a procedure removes the ghosts and indicates the equivalence of $E\Lambda$ and CG, it nevertheless leaves subtleties that have yet to be answered. 

Maldacena's boundary conditions are peculiar. In contrast to the full general solution that one would find for the tensor modes in the de Sitter background, by applying them to the perturbations, one obtains just a particular solution. This leads to a natural question  -- \textit{Is it possible to find boundary conditions that would keep the solution of the tensor modes in their full generality? }

Moreover, CG is a curious theory of gravity. Due to its invariance under conformal transformations, it is unable to distinguish between various conformally flat spacetimes. Thus, even if one would find perturbations that match those of the dS space in flat coordinates, one would not be able to determine if the resulting theory is the Minkowski spacetime with the dS perturbations or any other background that is connected to it via conformal transformations. In other words, at this point, it is unclear -- \textit{What are the conditions that determine if the background is de Sitter or Minkowski, and what are their consequences for the perturbations?}

In this paper, we will address these questions, by studying the interplay of different conditions that match the perturbation theory of Einstein and Conformal gravities for dS and Minkowski backgrounds. 

First, we will study the basics of CG, reviewing its degrees of freedom, and Maldacena's argument for its equivalence with ordinary gravity. Then, we will find the conditions that set the background equal to the de Sitter Universe. Along the lines of Maldacena, we will assume the Neumann Boundary Condition, and derive an additional one. In contrast to the condition of positive-frequency modes, we will find that this set of boundary conditions recovers the general solution for the perturbations which matches the perturbations around dS in $E\Lambda$.

This analysis will be followed by the study of perturbations of CG in the Minkowski spacetime. Surprisingly, we will find that if one wishes to recover a general solution for the tensor modes in this case, one has to abandon the Neumann boundary conditions entirely. Instead, we will derive the alternative boundary conditions, which will remove the ghost from the flat background and recover a general solution that matches with perturbations of Einstein's gravity. With this, we will remove the ghost of CG in de Sitter and Minkowski backgrounds on a classical level, recovering fully general solutions for the surviving modes.

\section{ {\textcolor{Black}{\Large \textbf{\textsc{Conformal Gravity -- A First Look}}}}}

In this chapter, we will review the main properties of conformal gravity (CG), necessary for the remainder of the paper. The action of conformal gravity is given by\footnote{ We use the signature $(-,+,+,+)$.} 
\begin{equation}\label{CGaction}
\begin{split}
S_{CG} & = \alpha_{CG}\int d^4x \sqrt{-g}W_{\lambda\mu\nu\rho}W^{\lambda\mu\nu\rho}\\
   & = \alpha_{CG}\int d^4x \sqrt{-g}\left(R_{\lambda\mu\nu\rho}R^{\lambda\mu\nu\rho}-2R_{\mu\nu}R^{\mu\nu}+\frac{1}{3}R^2\right).
\end{split} 
\end{equation}
Here, $a_{CG}$ is a dimensionless coupling, and $W_{\lambda\mu\nu\rho}$ is the Weyl tensor, given by
\begin{equation}
\begin{split}
      W_{\mu\nu\rho\sigma}&=R_{\mu\nu\rho\sigma}+\frac{1}{6}R(g_{\mu\rho}g_{\nu\sigma}-g_{\mu\sigma}g_{\nu\rho})\\
      &-\frac{1}{2}(g_{\mu\rho}R_{\nu\sigma}-g_{\mu\sigma}R_{\nu\rho}-g_{\nu\rho}R_{\mu\sigma}+g_{\nu\sigma}R_{\mu\rho}).
\end{split}
\end{equation}
Using the Gauss-Bonnet (GB) term,
\begin{equation}\label{GB}
 S_{GB}=\int d^4x\sqrt{-g}(R_{\mu\nu\rho\sigma}R^{\mu\nu\rho\sigma}-4R_{\mu\nu}R^{\mu\nu}+R^2),
\end{equation}
 the above action can be written as 
\begin{equation}\label{Sconfgrav}
\begin{split}
S_{CG} & = 2\alpha_{CG}\int d^4x \sqrt{-g}\left(R_{\mu\nu}R^{\mu\nu}-\frac{1}{3}R^2\right) + \alpha_{CG}S_{GB}.
\end{split} 
\end{equation}

The GB term is a total derivative, and thus does not contribute to the equations of motion: 
 \begin{equation}
    B_{\mu\nu}=0,
\end{equation}
where $B_{\mu\nu}$ the \textit{Bach tensor}:
\begin{equation}\label{Bach}
    \begin{split}
        B_{\mu\nu}=&\nabla_{\alpha}\nabla_{\mu} R^{\alpha}_{\;\nu}+\nabla_{\alpha}\nabla_{\nu} R^{\alpha}_{\;\mu}-\Box R_{\mu\nu}+\frac{1}{6}g_{\mu\nu}\Box R-\frac{1}{3}\left(\nabla_{\mu}\nabla_{\nu}R+\nabla_{\mu}\nabla_{\nu}R\right)\\
        &-2R_{\alpha\mu}R^{\alpha}_{\;\nu}+\frac{2}{3}RR_{\mu\nu}+\frac{1}{2}g_{\mu\nu}\left(R_{\alpha\beta}R^{\alpha\beta}-\frac{1}{3}R^2\right),
    \end{split}
\end{equation}
and $\Box=\nabla_{\alpha}\nabla^{\alpha}$. 
However, its presence is nevertheless important -- without it, the action (\ref{Sconfgrav}) would not be invariant under conformal transformations:
\begin{equation}
    \begin{split}
         &g_{\mu\nu}\to\Tilde{g}_{\mu\nu}=\Omega^2(x) g_{\mu\nu}\\
        &x^{\mu}\to\Tilde{x}^{\mu}=x^{\mu}.
    \end{split}
\end{equation}

In particular, one can find that upon this transformation the GB action becomes:
\begin{equation}
 \begin{split}
     S_{GB}\qquad\to\qquad\Tilde{S}_{GB}=S_{GB}+S_{\textit{rem}},
 \end{split}
\end{equation}
where 
\begin{equation}
\begin{split}
    S_{\textit{rem}}=-8\int d^4x\frac{\sqrt{-g}}{\Omega^2}&\left[R^{\mu\nu}\nabla_{\mu}\Omega\nabla_{\nu}\Omega-\Box \Omega\Box \Omega+\nabla_{\mu}\nabla_{\nu}\Omega\nabla^{\mu}\nabla^{\nu}\Omega\right.\\ &\left.+\frac{2}{\Omega}\left(\nabla^{\nu}\Omega\nabla_{\nu}\Omega\Box\Omega-\nabla_{\mu}\nabla_{\nu}\Omega\nabla^{\mu}\Omega\nabla^{\nu}\Omega\right)\right].
\end{split}
\end{equation}
Here, to simplify the expression, we have used the contracted Bianchi identity. The remaining contribution of (\ref{Sconfgrav}) precisely cancels this contribution so that the overall action is invariant. 

CG has numerous classical solutions  \cite{Riegert:1984zz, Lu:2012xu, Mannheim:1988dj, Corral:2021xsu, Dunajski:2013zta, Wang:2022err, Dzhunushaliev:2021cgu, Liu:2012xn}. A particularly interesting class of these are Einstein spaces -- solutions for which the Ricci tensor is proportional to the metric. At the same time, these solutions are solutions of both $E\Lambda$ and CG.  It has been shown that when evaluated on these spaces, the action of the CG equals the renormalized Einstein-AdS gravity \cite{Anastasiou:2016jix}. In this paper, we will consider Minkowski and dS solutions in CG, focusing on the perturbation theory around these backgrounds.

%%%%%%%%%%%%%%%%%%%%%%%%%%%%%%%%%%%%%%%%%%%%%%%%%%%%%%%%%%%%%%%%%%%%%%%%%%%
\subsection{ {\textcolor{Black}{\large
 \textbf{\textsc{The Degrees of Freedom}}}}}
\normalsize

Let us now examine the \textit{degrees of freedom (dof)} of CG. Due to the invariance under conformal transformations, CG cannot distinguish between different conformally flat spacetimes. Thus, even at the level of the linearised theory, the actions for de Sitter Universe and Minkowski spacetime will not differ. In order to see this, let us expand the metric as 
\begin{equation}
    g_{\mu\nu}=a^2(\eta)\left(\eta_{\mu\nu}+h_{\mu\nu}\right),
\end{equation}
where $a(\eta)$ is the scale factor, which is  for dS Universe given by
\begin{equation}
   a(\eta)=-\frac{1}{H_{\Lambda}\eta},
\end{equation}
where $H_{\Lambda}$ is the Hubble constant. For the Minkowski spacetime, $a(\eta)$ equals unity. Then, to the lowest order in metric perturbations, the action becomes: 
\begin{equation}
   \begin{split}
        S_{CG} = 2\alpha_{CG}\int d^4x&\left(\frac{1}{6}\partial_{\mu}\partial_{\nu}h^{\mu\nu}\partial_{\alpha}\partial_{\beta}h^{\alpha\beta}-\frac{1}{2}\partial_{\nu}\partial_{\alpha}h^{\alpha\beta}\partial^{\mu}\partial^{\nu}h_{\mu\beta}\right.\\
        &\left.+\frac{1}{4}\Box h_{\alpha\beta}\Box h^{\alpha\beta}+\frac{1}{6}\partial^{\mu}\partial^{\nu}h_{\mu\nu}\Box h-\frac{1}{12}\Box h\Box h\right).
   \end{split}
\end{equation}
Here, we are raising and lowering the indices with the Minkowski metric, $h=h^{\mu}_{\mu}$ denotes the trace of the perturbations, and the partial derivatives are denoted by $\partial_{\mu}=\frac{\partial}{\partial x^{\mu}}$. 

We can notice that the scale factor has entirely dropped out from the action. As a result, on the level of action, CG does not distinguish between the Minkowski and dS spacetime. In order to find the \textit{dof}, let us now decompose the perturbations according to the irreducible representations of the SO(2) group: 
\begin{equation*}\label{decomposition}
    \begin{split}
        &h_{00}=2\phi\\
        &h_{0i}=B_{,i}+S_i,\qquad\qquad S_{i,i}=0\\
        &h_{ij}=2\psi\delta_{ij}+2E_{,ij}+F_{i,j}+F_{j,i}+h_{ij}^{T},\qquad\qquad F_{i,i}=0,\quad h_{ij,i}^{T}=0,\quad h_{ii}^{T}=0,
    \end{split}
\end{equation*}
where $,i=\frac{\partial}{\partial x^i}$. Then, the Lagrangian density becomes
\begin{equation}
    \mathcal{L}=\mathcal{L}_S+\mathcal{L}_V+\mathcal{L}_T,
\end{equation}
where:
\begin{equation}
    \begin{split}
        \mathcal{L}_S=&\frac{4}{3}\alpha_{CG}\left(\phi\Delta^2\phi+2\phi\Delta^2\psi+\psi\Delta^2\psi- B\Delta^2\Ddot{B}+\Ddot{E}\Delta^2\Ddot{E}\right.\\&\left.-2\phi\Delta^2\Dot{B}+2\phi\Delta^2\Ddot{E}+2\psi\Delta^2\Ddot{E}-2\psi\Delta^2\Dot{B}-2\Ddot{E}\Delta^2\Dot{B}\right)\\\\
        \mathcal{L}_V=&\alpha_{CG}\left(-\Ddot{F}_i\Delta\Ddot{F}_i+\Ddot{F}_i\Delta^2F_i+\Ddot{S}_i\Delta S_i-S_i\Delta^2S_i-2\Ddot{S}_i\Delta\Dot{F}_i-2\Dot{S}_i\Delta^2F_i\right)\\\\
        \mathcal{L}_T=&\frac{\alpha_{CG}}{2}\Box h_{ij}^T\Box h_{ij}^T
    \end{split}
\end{equation}
Here, the dot denotes a derivative with respect to the (conformal) time $\eta$. Let us now analyze the perturbations separately. 

\textcolor{YellowOrange}{$\diamond$}\;\;\textsc{\textbf{Scalar Perturbations}}

We can notice that the scalars $\phi$ and $\psi$ are not propagating -- the second time derivatives in these fields are missing. By varying the action with respect to $\phi$, we obtain the following constraint: 
\begin{equation}\label{phiconstraint}
    \phi=\dot{B}-\Ddot{E}-\psi.
\end{equation}
Substituting it back into the action, we can see that the Lagrangian density corresponding to the scalar perturbations vanishes entirely. This leads us to the conclusion that there are no scalar \textit{\textit{dof}} in CG. 

\textcolor{YellowOrange}{$\diamond$}\;\;\textsc{\textbf{Vector Perturbations}}

In order to study the vector perturbations, we will define 
\begin{equation}
    V_i=S_i-\dot{F}_i. 
\end{equation}
Then, the Lagrangian density becomes only a function of one vector mode
\begin{equation}\label{vmodes}
    \mathcal{L}_V=\alpha_{CG}\Delta V_i\Box V_i. 
\end{equation}
Its equation of motion is given by
\begin{equation}\label{vmodeseq}
    \Box V_i=0,
\end{equation}
leading us to two \textit{dof} that arise from the vector perturbations. 

\textcolor{YellowOrange}{$\diamond$}\;\;\textsc{\textbf{Tensor Perturbations}}

Finally, the tensor perturbations satisfy: 
\begin{equation}\label{CGtensor}
    \Box^2h_{ij}^T=0
\end{equation}
This equation of motion gives us an additional four \textit{dof}, two of which are the ghost \textit{dof}. Thus, in total, CG has 6 \textit{dof} -- 2 vector and four tensor ones. 

One should notice that the above finding are independent of the choice of the coordinate system. Under infinitesimal coordinate transformations 
\begin{equation}
    x^{\mu}\to \Tilde{x}^{\mu}=x^{\mu}+\xi^{\mu}
\end{equation}
the metric perturbation transforms as 
\begin{equation}
    h_{\mu\nu}\to\Tilde{h}_{\mu\nu}=h_{\mu\nu}-\xi_{\mu,\nu}-\xi_{\nu,\mu}
\end{equation}
Decomposing
\begin{equation}
\xi_{\mu}=(\xi_0, \xi_i),\qquad \xi_i=\xi_i^T+\zeta_{,i},\qquad  \xi_{i,i}^T=0
\end{equation}
we then find
\begin{equation}
    \begin{split}
        &\phi\to\Tilde{\phi}=\phi-\dot{\xi}_0\qquad \psi\to\Tilde{\psi}=\psi\qquad
        B\to\Tilde{B}=B-\xi_0-\dot{\zeta}\qquad E\to\Tilde{E}=E-\zeta\\\\
        &S_i\to\Tilde{S}_i=S_i-\dot{\xi}_i^T\qquad F_i\to\Tilde{F}_i=F_i-\xi_i^T\\\\
        & h_{ij}^{T}\to\Tilde{h}_{ij}^{T}=h_{ij}^{T}
    \end{split}
\end{equation}
Thus, the tensor perturbations $h_{ij}^{T}$, vector perturbations $V_i$ and the scalar one $\psi$ are all gauge invariant quantities \cite{Cosmo}. In addition, upon the infinitesimal coordinate transformation, the constraint that is obeyed by the scalars (\ref{phiconstraint}) will not change its form. As a result, the above findings for the \textit{dof} of CG will be independent of the choice of the reference frame.  

\subsection{  {\textcolor{Black}{
 \textbf{\textsc{Maldacena's Boundary Conditions}}}}}
\normalsize

Clearly, the presence of ghosts is not ideal to make the case for Conformal Gravity. However, in \cite{Maldacena:2011mk} it was argued that they could be avoided by choosing the Neumann Boundary and Positive Frequency Mode conditions. In this subsection, we will summarize Maldacena's argument and demonstrate how the solutions of the tensor modes recombine to form a particular solution of the de Sitter tensor mode, while the same choice sets the vector modes to zero. 

Following \cite{Maldacena:2011mk}, as a first step let us choose the synchronous gauge:
\begin{equation}
    h_{0\alpha}=0
\end{equation}
that sets $S_i=0$.  In Fourier space, the vector modes can be expressed as 
\begin{equation}
     V_i=\int \frac{d^3k}{(2\pi)^{3/2}}v^{}_k\varepsilon_{i}^Te^{i\Vec{k}\Vec{x}}
\end{equation}
where $\varepsilon_i$ is the polarisation tensor. Then, the solution of (\ref{vmodeseq}) is given by
\begin{equation}\label{vmodes}
    v_k=A_Ve^{ik\eta}+B_Ve^{-ik\eta}
\end{equation}
where $A_V$ and $B_V$ are constants of integration.

Similarly, writing the tensor modes also in the Fourier space, 
\begin{equation}
     h^{T}_{ij}=\int \frac{d^3k}{(2\pi)^{3/2}}h^{}_k\varepsilon_{ij}^Te^{i\Vec{k}\Vec{x}},
\end{equation}
where $\varepsilon_{ij}^T$ are the polarisation tensors, we find that the general solution of (\ref{CGtensor}) is given by 
\begin{equation}
    h_k=A_Te^{ik\eta}+B_Te^{-ik\eta}+\eta\left[C_Te^{ik\eta}+D_Te^{-ik\eta}\right]
\end{equation}
where $A_T,B_T,C_T$ and $D_T$ are constants of integration. Here, we see that the first two terms correspond to the healthy tensor mode. The second two are linear in time, and as such are the source of the linear instability \cite{Boulware:1983td} -- they are known as the ghosts. 

\textsc{The Neumann Boundary Condition} is a statement that 
\begin{equation}
    \left.\dot{g}_{ij}\right|_{\eta=0}=0
\end{equation}
For the tensor modes, this implies that 
\begin{equation}
    \left.\dot{h}^T_{ij}\right|_{\eta=0}=0 
\end{equation}
and, for vector modes: 
\begin{equation}
    \left.\dot{F}_{i}\right|_{\eta=0}=0.
\end{equation}
Thus, for the gauge-invariant vector modes, this becomes the Dirichlet condition:
\begin{equation}
    \left.V_{i}\right|_{\eta=0}=0.
\end{equation}

\textsc{The Positive Frequency Mode Condition} is the requirement that at very early times only the positive-frequency modes are present. In other words, we have
\begin{equation}
    \lim_{\eta\to-\infty} h_k\sim e^{-ik\eta}\qquad \text{and}\qquad 
    \lim_{\eta\to-\infty}v_k\sim e^{-ik\eta}.
\end{equation}

When applied to the vector modes, the two conditions make them vanish entirely.

The solutions of the tensor modes, on the other hand, recombine such that only one particular solution remains: 
\begin{equation}
    h_k=B_T\left(1+ikt\right)e^{-ikt}. 
\end{equation}
This is precisely the form of one of the tensor mode solutions of Einstein gravity with the de Sitter background in flat coordinates, that satisfy \cite{Cosmo}: 
\begin{equation}\label{dSsol}
    h^{T(dS)}_{ij}=\int \frac{d^3k}{(2\pi)^{3/2}}h^{(dS)}_k\varepsilon_{ij}^Te^{i\Vec{k}\Vec{x}}\qquad\qquad h^{(dS)}_k=C_{\pm}\left(1\pm ik\eta\right)e^{\mp ik\eta},
\end{equation}
where $C_{\pm}$ are constants of integration.

Thus, it seems we have simply recovered the perturbation theory in the de Sitter space. Nevertheless, as we have pointed out in the introduction, this result hides subtleties that are yet to be answered. 
Due to the invariance under conformal transformations, conformal gravity cannot distinguish between two spacetimes, that are related by a conformal transformation. Thus, one possibility to write the final result would be 
\begin{equation}
ds^2=\eta_{\mu\nu}dx^{\mu}dx^{\nu}+h_{ij}^Tdx^idx^j,
\end{equation}
while another would be 
\begin{equation}
ds^2=\frac{1}{H_{\Lambda}^2\eta^2}\left(\eta_{\mu\nu}dx^{\mu}dx^{\nu}+h_{ij}^Tdx^idx^j\right).
\end{equation}
At this point, the background is not apriori-determined. As a result, one could end up with the metric whose background is Minkowski spacetime, while the perturbations behave as if they were around the de Sitter background -- a result that does not seem to be realistic but within this framework could be possible. 
Furthermore, the solution that we have obtained is only a particular one, leading us to the natural question if is it possible to obtain a more general solution, just like one can obtain in the case of de Sitter Universe. 
In the next chapter, we will study these questions, finding the conditions that select the de Sitter background and leading to the general solution for the tensor perturbations.

%%%%%%%%%%%%%%%%%%%%%%%%%%%%%%%%%%%%%%%%%%%%%%%%%%%%%%%%%%%%%%%%%%%%%%%%%%%%%%%%%%%%%%%%%%%

\section{{\textcolor{Black}{\Large \textbf{\textsc{de Sitter Selection}}}}}

Previously, we have seen that imposing Maldacena's boundary conditions leads us to an ambiguity -- \textit{Which condition determines the background space-time to be de Sitter?} To answer it, let's consider how can one obtain Einstein's gravity with a cosmological constant ($E\Lambda$), starting from the CG action. 

In terms of the traceless part of the Ricci tensor, 
\begin{equation}
    R_{\mu\nu}=H_{\mu\nu}-\frac{g_{\mu\nu}}{4}R,
\end{equation}
the action of conformal gravity can be written as 
\begin{equation}\label{CG}
    \begin{split}
        S_{CG}=\alpha_{CG}\int d^4x \sqrt{-g}\left[2H_{\mu\nu}H^{\mu\nu}-\frac{4}{3}\Lambda\left(R-2\Lambda\right)-24\left(\frac{R}{12}-\frac{\Lambda}{3}\right)^2\right]+\alpha_{CG}S_{GB},
    \end{split}
\end{equation}
where $\Lambda=3H_{\Lambda}^2$ is the cosmological constant. We can notice that the middle term looks like $E\Lambda$, while the remaining terms set the solutions of CG apart from the ordinary gravity. (Up to the GB term that we will comment on later on. )
Thus, to recover $E\Lambda$, we should set 
\begin{equation}\label{dSselection}
    H_{\mu\nu}=0\qquad \text{and}\qquad R=12H_{\Lambda}^2.
\end{equation}

\subsection{  {\textcolor{Black}{
 \textbf{\textsc{The Starobinsky Expansion}}}}}
\normalsize

By studying the inhomogeneous asymptotic structure of an expanding Universe filled with an effective cosmological constant, Starobinsky in \cite{Starobinsky} has introduced an asymptotic expansion of the metric at $t\to\infty$, known as \textit{the Starobinsky expansion}, which is given by:
\begin{equation}
    ds^2=-dt^2+\gamma_{ij}(\Vec{x})dx^idx^j,\qquad \gamma_{ij}=e^{2H_{\Lambda}t}a_{ij}+b_{ij}+e^{-H_{\Lambda}t}c_{ij}+...
\end{equation}
In terms of the conformal time, 
\begin{equation}
    a(\eta)d\eta=dt\qquad \text{where} \qquad a(\eta)=-\frac{1}{H_{\Lambda}\eta}, \qquad\text{and}\qquad 3H^2_{\Lambda}=\Lambda,
\end{equation}
the above expansion can be written as 
\begin{equation}\label{StarExp}
    ds^2=\frac{1}{H_{\Lambda}^2\eta^2}\left[-d\eta^2+\sum_{n=0}^{\infty}\frac{(-1)^n}{n!}(\eta H_{\Lambda})^ng^{(n)}_{ij}(\Vec{x})dx^idx^j\right]
\end{equation}
with 
\begin{equation}\label{NBC}
    g^{(1)}_{ij}=0. 
\end{equation}
Here, $g^{(n)}_{ij}(\Vec{x})$ are coefficients that depend only on space. 
As we have seen, in CG, one can equivalently remove the overall scale factor in (\ref{StarExp}) with which this expansion would become an expansion of the metric at $\eta=0$. As a result, the above coefficient would simply become the derivatives of the metric, and thus, the Neumann Boundary Condition (NBC) would simply become the statement given in the equation (\ref{NBC}). 

In this subsection, we will use the Starobinsky expansion and the conditions (\ref{dSselection}) in order to find relations satisfied by metric coefficients $g_{ij}^{(n)}$. This procedure will be similar to that of \cite{Anastasiou:2020mik} where the relation between metric coefficients was found by using the Fefferman-Graham expansion \cite{FGex,Graham:1999jg} in AdS. Knowing the relation between different metric coefficients will then help us to determine the boundary conditions that the metric should satisfy such that its perturbations recover the full general solution of the dS space.
By working with the expansion that contains a general scale factor
\begin{equation}\label{SExp}
    ds^2=a^2(\eta)\left[-d\eta^2+\sum_{n=0}^{\infty}\frac{(-1)^n}{n!}(\eta H_{\Lambda})^ng^{(n)}_{ij}(\Vec{x})dx^idx^j\right], \qquad g_{ij}^{(1)}=0
\end{equation}
we will show that the only possible background that fulfills the conditions (\ref{dSselection}) is the de Sitter Universe. 

By substituting (\ref{SExp}) into the Ricci scalar and the traceless part of the Ricci tensor, we find: 
\begin{equation}
   \begin{split}
        R&=\frac{1}{a^2}\left[6\frac{\ddot{a}}{a}+R^{(0)}+H_{\Lambda}^2g^{(2)}\left(3\eta\frac{\dot{a}}{a}+1\right)-H_{\Lambda}^3g^{(3)}\left(\eta+\frac{3}{2}\eta^2\frac{\dot{a}}{a}\right)+\mathcal{O}(\eta^2)\right]
\\\\
        H_{00}&=3\frac{(\dot{a})^2}{a^2}-\frac{3}{2}\frac{\ddot{a}}{a}+\frac{1}{4}R^{(0)}+\frac{H_{\Lambda}^2}{4}g^{(2)}\left(\eta\frac{\dot{a}}{a}-1\right)+H_{\Lambda}^3\eta g^{(3)}\left(\frac{1}{4}-\frac{\dot{a}}{8a}\eta\right) +\mathcal{O}(\eta^2)\\\\
          H_{0i}&=\frac{H_{\Lambda}^2}{2}\eta\left(D_jg_i^{j(2)}-D_ig^{(2)}\right)+\mathcal{O}(\eta^2)
    \end{split}
\end{equation}
and
\begin{equation}
    \begin{split}
        H_{ij}&=R_{ij}^{(0)}-\frac{1}{4}R^{(0)}g_{ij}^{(0)}+g_{ij}^{(0)}\left(\frac{(\dot{a})^2}{a^2}-\frac{\ddot{a}}{2a}\right)\\&+H_{\Lambda}^2\left[\frac{1}{2}g_{ij}^{(2)}+\frac{1}{2}\frac{(\dot{a})^2}{a^2}\eta^2g_{ij}^{(2)}+\frac{\dot{a}}{a}\eta g_{ij}^{(2)}-\frac{1}{4}\frac{\ddot{a}}{a}\eta^2g_{ij}^{(2)}-\frac{g_{ij}^{(0)}}{4}g^{(2)}\left(1+\frac{\dot{a}}{a}\eta\right)\right]\\
        &-H_{\Lambda}^3\left[g_{ij}^{(3)}\left(\frac{\eta}{2}-\frac{\ddot{a}\eta^3}{12 a}+\frac{(\dot{a})^2\eta^3}{6a^2}+\frac{\dot{a}\eta^2}{2a}\right)-g_{ij}^{(0)}g^{(3)}\frac{\eta}{4}\left(1+\frac{a\eta}{2a}\right)\right]
    \end{split}
\end{equation}
Here, the curvature terms $R_{ij}^{(0)}$, $R^{(0)}$ and covariant derivatives $D_i$ correspond to the three-dimensional metric $g^{(0)}_{ij}$. Furthermore, in these expressions, we raise and lower the indices with $g^{(0)}_{ij}$, while $g^{(n)}=g^{(0)ij}g_{ij}^{(n)}$ denotes the trace of the metric. 

The above equations should be identified with the conditions (\ref{dSselection}), and solved at each power of the conformal time. 
The Ricci scalar, and the $H_{00}$ component imply that 
\begin{equation}\label{scalefactor}
    \frac{(\dot{a})^2}{a^2}-\frac{\ddot{a}}{2a}=0\qquad \frac{\ddot{a}}{a^3}=2H_{\Lambda}^2
\end{equation}
The only solution to these equations is the scale factor of the de Sitter Universe: 
\begin{equation}
  a=-\frac{1}{H_{\Lambda}\eta}. 
\end{equation}
This means that the background is now determined. Imposing the conditions (\ref{dSselection}), we break the conformal invariance and select the particular space-time background. 

We have seen that the NBC is implemented in the Starobinsky expansion and plays a role in selecting the solutions of the metric perturbations. Let us now derive another boundary condition that will replace the requirement of the positive frequency modes, and recover the general solution of the perturbations. 

Considering the higher-order terms, the Ricci scalar and the $H_{00}$ component imply: 
\begin{equation}\label{CGC1}
    g^{(2)}=\frac{R^{(0)}}{2H_{\Lambda}^2}
\end{equation}
and
\begin{equation}\label{CGC3}
    g^{(3)}=0
\end{equation}
Substituting this into $H_{ij}=0$, we then find
\begin{equation}\label{CGC2}
    g^{(2)}_{ij}=\frac{2}{H_{\Lambda}^2}\left(R_{ij}^{(0)}-\frac{1}{4}g_{ij}^{(0)}R^{(0)}\right).
\end{equation}
Finally, the $H_{0i}$ component gives us 
\begin{equation}\label{CGC4}
    D_jg_i^{j(2)}=D_ig^{(2)}.
\end{equation}

These conditions together with (\ref{scalefactor}) and NBC guarantee us that we will obtain the de Sitter Universe together with appropriate perturbations. To see this, let us apply them to the metric perturbations that we have previously discussed.

\subsection{  {\textcolor{Black}{
 \textbf{\textsc{CG and the dS Perturbations}}}}}
\normalsize
In the previous part, we have derived the conditions that metric coefficients should satisfy such that the traceless part of the Ricci tensor vanishes, while the Ricci scalar takes the values as if it would in the de Sitter space. The key to making these requirement lies in the action of CG -- with such values, the pure contribution of CG would vanish, and one would only obtain a term resembling the $E\Lambda$, together with the GB term. 

By assuming the general value of the scale factor, we have shown that then, the only possible background is the de Sitter space. However, the question of the perturbations around this space still remains. In this subsection, we will show that the equivalence of CG and $E\Lambda$ extends even further -- on the level of perturbations.

By decomposing the metric as 
\begin{equation}\label{Starexpansion}
    g_{\mu\nu}=a^2(\eta)\left(\eta_{\mu\nu}+h_{\mu\nu}\right),
\end{equation}
we can identify the metric coefficients of the Starobinsky expansion $g^{(n)}_{\mu\nu}$ and the metric perturbations $h_{\mu\nu}$ as
\begin{equation}
    g_{\mu\nu}^{(0)}=\left.h_{\mu\nu}\right|_{\eta=0}\qquad  -g_{\mu\nu}^{(1)}=\left.\dot{h}_{\mu\nu}\right|_{\eta=0}\qquad H_{\Lambda}^2g_{\mu\nu}^{(2)}=\left.\ddot{h}_{\mu\nu}\right|_{\eta=0}
\end{equation}
With this connection, NBC implies: 
\begin{equation}
  \left.\dot{h}_{ij}\right|_{\eta=0}=0
\end{equation}
We can notice that the condition on the second-order coefficients (\ref{CGC2}) involves the 3-curvatures. In order to find how this expression translates to the metric perturbations, we linearize the Ricci tensor and scalar and find:  
\begin{equation}
    R^{(0)}_{ij}=\left.\frac{1}{2}\left[\partial_k\partial_ih_j^k+\partial_k\partial_jh_i^k-\Delta h_{ij}-\partial_j\partial_i h_k^k\right]\right|_{\eta=0},
\end{equation}
and
\begin{equation}
    R^{(0)}=\left.\partial_i\partial_jh^{ij}-\Delta h_i^i\right|_{\eta=0}.
\end{equation}
Here, the indices are raised and lowered with the Minkowski metric. With this, (\ref{CGC2}) becomes: 
\begin{equation}\label{SeconCond}
   \left.\Ddot{h}_{ij}\right|_{\eta=0}=2\left(R_{ij}^{(0)}-\frac{1}{4}g_{ij}^{(0)}R^{(0)}\right).
\end{equation}
Lastly, the conditions (\ref{CGC3}) and (\ref{CGC4}) become respectively 
\begin{equation}\label{remcond}
    \left.\frac{\partial^3 h^i_i}{\partial \eta^3} \right|_{\eta=0}=0\qquad\text{and}\qquad \left.\partial_j\ddot{h}_{i}^j\right|_{\eta=0}=\left.\partial_i \Ddot{h}^j_j\right|_{\eta=0}.
\end{equation}
Let us now consider the vector and tensor modes separately. 

\textcolor{YellowOrange}{$\diamond$}\;\;\textsc{\textbf{Tensor Perturbations}}

Previously, we have found the following solutions for the tensor modes: 
\begin{equation}\label{genSolutionTens}
     h^{T}_{ij}=\int \frac{d^3k}{(2\pi)^{3/2}}h^{}_k\varepsilon_{ij}^Te^{i\Vec{k}\Vec{x}},
\end{equation}
where
\begin{equation}
    h_k=A_Te^{ik\eta}+B_Te^{-ik\eta}+\eta\left[C_Te^{ik\eta}+D_Te^{-ik\eta}\right].
\end{equation}
Applying the NBC, we find 
\begin{equation}
    ik(A_T-B_T)+C_T+D_T=0
\end{equation}
In Fourier space, the second condition (\ref{SeconCond}) becomes
\begin{equation}\label{dSNBCcond}
    \left.\ddot{h}_k\right|_{\eta=0}=\left.k^2h_k\right|_{\eta=0}
\end{equation}
from which we obtain 
\begin{equation}
    ik(A_T+B_T)+C_T-D_T=0
\end{equation}
Solving the two equations for the constants, we find
\begin{equation}
    C=-ikA\qquad \text{and}\qquad D=ikB
\end{equation}
and thus we find a general solution: 
\begin{equation}
    h_k(\eta)=A(1-ik\eta)e^{ik\eta}+B(1+ik\eta)e^{-ik\eta},
\end{equation}
which agrees with the de Sitter solution in Einstein's gravity for the tensor modes. As these modes are traceless and transverse, the remaining two conditions (\ref{remcond}) will be identically satisfied. 

The above values of the tensor modes were found for the re-scaled perturbations, $h_{\mu\nu}$, the \textit{true} perturbations around the de Sitter background
\begin{equation}
    g_{\mu\nu}=g_{\mu\nu}^{(0)}+\delta g_{\mu\nu},\qquad \text{where}\qquad g_{\mu\nu}^{(0)}=a^2(\eta)\eta_{\mu\nu},
\end{equation}
are then given by: 
\begin{equation}
     \delta g_{ij}^T(\eta)=a^2(\eta)h_{ij}^T=\int \frac{d^3k}{(2\pi)^{3/2}}\frac{1}{H_{\Lambda}^2\eta^2}\left[A(1-ik\eta)e^{ik\eta}+B(1+ik\eta)e^{-ik\eta}\right]\varepsilon_{ij}^Te^{i\Vec{k}\Vec{x}}
\end{equation}

\textcolor{YellowOrange}{$\diamond$}\;\;\textsc{\textbf{Vector Perturbations}}

In contrast to the tensor modes, the gauge-invariant vector mode vanishes under the derived conditions. For them, the NBC becomes the Dirichlet condition: 
\begin{equation}
    \left.V_i\right|_{\eta=0}=0,
\end{equation}
while the second condition (\ref{SeconCond}) gives us Neumann condition:
\begin{equation}
    \left.\dot{V}_i\right|_{\eta=0}=0
\end{equation}
Thus, for a general solution that we have previously found: 
\begin{equation}
    v_k=A_Ve^{ik\eta}+B_Ve^{-ik\eta}
\end{equation}
these two imply 
\begin{equation}
    A=B=0. 
\end{equation}

Thus, in this section, we have seen that the ghost \textit{dof} completely disappears from the CG, aligning its perturbations with those of $E\Lambda$ in the de Sitter case. Moreover, by performing the analysis with a general factor, we have seen that the de Sitter Universe is the only possible background under the conditions given by (\ref{dSselection}). One might wonder how out of all possible conformally flat space-times, we have ended up with this particular one. The reason is that the conditions (\ref{dSselection}) break conformal invariance. For example, under conformal transformations, the Ricci scalar transforms as 
\begin{equation}
    R\qquad\to\qquad\Tilde{R}=\frac{R}{\Omega^2}-6\frac{\nabla^{\mu}\nabla_{\mu}\Omega}{\Omega^3}.
\end{equation}
Due to this, the arbitrariness of the scale factor is lost, and the only possibility that remains is the de Sitter Universe.

\subsection{  {\textcolor{Black}{
 \textbf{\textsc{Can we Ignore the Bauss-Bonnet Term?}}}}}
\normalsize

Up to this point, we were aiming to set the first and last term of the action 
(\ref{CG}) to zero, in order to recover $E\Lambda$. The GB term, which is also present in the action is a total derivative, and will thus not affect the equations of motion. Nevertheless, due to the following argument, we will see that in the dS case, the CG will not only reduce to $E\Lambda$, but the GB term should also remain in the final action. 

Let us consider the action (\ref{CG}) and substitute the values of the curvatures for the de Sitter space. In this case, the following two terms vanish: 
\begin{equation}
    H_{\mu\nu}H^{\mu\nu}=0\qquad \qquad \left(\frac{R}{12}-\frac{\Lambda}{3}\right)^2=0
\end{equation}
The Gauss-Bonnet term, on the other hand, gives a non-vanishing contribution: 
\begin{equation}
    \alpha_{CG}S_{GB}=\frac{8}{3}\alpha_{CG}\Lambda^2\int d^4x\sqrt{-g}
\end{equation}
and the second term, which corresponds to the Einstein action with the cosmological constant gives an opposite contribution:
\begin{equation}
   -\alpha_{CG}\frac{4}{3}\Lambda\left(R-2\Lambda\right)=-\alpha_{CG}\frac{8}{3}\Lambda^2
\end{equation}
which cancels the Gauss-Bonnet term. It is not surprising that the overall action is zero -- dS space is conformally flat, so the corresponding Weyl tensor is zero. The action of Einstein's gravity with a cosmological term gives however a non-vanishing contribution: 
\begin{equation}
    S_{EH}=\frac{M_{pl}^2}{2}\int d^4x\sqrt{-g}(R-2\Lambda)=M_{pl}^2\Lambda\int d^4x\sqrt{-g}
\end{equation}
Therefore, the only possible way to obtain Einstein's gravity from CG is to keep the Gauss-Bonnet term intact in the action in order to cancel the overall contribution of Einstein's Gravity. 

One should also note that the connection of the two theories relates the Planck mass with the cosmological constant and the CG coupling in the following way: 
\begin{equation}\label{MplLam}
    M_{pl}^2=-\frac{8}{3}\alpha_{CG}\Lambda.
\end{equation}
 For $\alpha_{CG}$ of the order of unity, the cosmological constant (and therefore the curvature) is of the same order as the Planck mass. At this point, we enter the regime of quantum gravity. Thus, in order to avoid it, we should require that the value of $\alpha_{CG}$ is much larger than one. 
We can roughly see that this corresponds to the weakly-coupled regime of conformal gravity. Expanding the metric, the action for the perturbations can be written as
\begin{equation}
    S_{CG}=\alpha_{CG}\int d^4x\left(\frac{h^2}{L^4}+\frac{h^3}{L^4}+...\right),
\end{equation}
where $\frac{1}{L}$ denotes a derivative, and $h$ denotes the perturbations. Canonically normalizing the tensor field with: 
\begin{equation}
    h_{ n}=\sqrt{\alpha_{CG}}h
\end{equation}
the quadratic term becomes 
\begin{equation}
    \alpha_{CG}h^2=h_{n}^2
\end{equation}
while for the cubic we have: 
\begin{equation}
    \alpha_{CG}h^3=\frac{h_n^3}{\sqrt{\alpha_{CG}}}
\end{equation}
As a result, we can see that the self-interacting term is smaller than the leading, kinetic term if 
\begin{equation}
    \sqrt{\alpha_{CG}}\gg 1
\end{equation}
which corresponds to the weakly coupled regime. 

\section{{\textcolor{Black}{\Large\textbf{\textsc{The ABC of the Minkowski Space}}}}}

In the previous chapter, we have found the conditions in which CG reduces to ordinary gravity with a cosmological constant and a GB term for the de Sitter space. These conditions break the conformal invariance, distinguishing between the Minkowski and de Sitter space -- property that was initially absent in the CG. As a result, the ghosts that were present in the CG recombine with the healthy modes and ultimately form the two healthy tensor modes. 

Here, we will in turn investigate the conditions that will recover the Minkowski space. Following the previous procedure, let us first consider the CG action, which we will now write in the following way: 
\begin{equation}\label{CGMInkowski}
    \begin{split}
        S_{CG}=\int d^4x \sqrt{-g}\left[2\alpha_{CG}H_{\mu\nu}H^{\mu\nu}-\frac{\alpha_{CG}}{2}R\left(\frac{R}{3}+\frac{M_{pl}^2}{\alpha_{CG}}\right)+\frac{M_{pl}^2}{2}R\right]+\alpha_{CG}S_{GB}
    \end{split}
\end{equation}

We can see that the third term now corresponds to Einstein's gravity, while the remaining ones are the contributions that arise from CG. Clearly, to recover the Minkowski space we have to set: 
\begin{equation}\label{MinkConditions}
    H_{\mu\nu}=0\qquad \text{and}\qquad R=0. 
\end{equation}
In the following, we will derive what boundary conditions are implied with these requirements, with a procedure similar to the previous chapter, and apply them to the metric perturbations. 

\subsection{  {\textcolor{Black}{
 \textbf{\textsc{Minkowski selection and the Neumann Boundary Conditions}}}}}
\normalsize
Following the line of the previous section, we will apply the Starobinsky expansion: 
\begin{equation}\label{StarExpM}
    ds^2=a^2(\eta)\left[-d\eta^2+\sum_{n=0}^{\infty}\frac{(-1)^n}{n!}\eta^ng^{(n)}_{ij}(\Vec{x})dx^idx^j\right]
\end{equation}
with 
\begin{equation}
    g_{ij}^{(1)}=0,
\end{equation}
to the conditions (\ref{MinkConditions}), and derive the resulting relations between the metric coefficients. As before, we will work with the general scale factor and show that (\ref{MinkConditions}) imply that the spacetime should now be flat. 

From the conditions
\begin{equation}
    R=0\qquad\text{and}\qquad H_{00}=0,
\end{equation}
we find
\begin{equation}
a''=0    \qquad\text{and}\qquad a'=0.
\end{equation}
Clearly, the scale factor is a constant which we can set to unity without the loss of generality. In addition, the same conditions imply: 
\begin{equation}
    g^{(2)}+R^{(0)}=0\qquad \text{and}\qquad   g^{(2)}-R^{(0)}=0.
\end{equation}
From here, it follows that 
\begin{equation}
    g^{(2)}=0\qquad\text{and}\qquad R^{(0)}=0. 
\end{equation}
Then, from 
\begin{equation}
    H_{ij}=0
\end{equation}
we find: 
\begin{equation}\label{NBCMink}
    g_{ij}^{(2)}=-2R_{ij}^{(0)}. 
\end{equation}
Finally, at $\mathcal{O}(\eta)$ we find: 
\begin{equation}
    g_{ij}^{(3)}=0. 
\end{equation}
Let us now apply these conditions to the perturbations. For vector perturbations, these conditions imply: 
\begin{equation}
    \left.V_i\right|_{\eta=0}=0\qquad\text{and}\qquad  \left. \dot{V}_i\right|_{\eta=0}=0.
\end{equation}
Thus, we find no vector modes -- both constants of integration are zero due to these boundary conditions.

For the tensor modes,  (\ref{NBCMink}) becomes:
\begin{equation}\label{MinkcondNBC}
    \left.\Ddot{h}_k\right|_{\eta=0}=\left.-k^2h_k\right|_{\eta=0}
\end{equation}
in the Fourier space. This condition relates two of the constants of the general solution (\ref{genSolutionTens}) as 
\begin{equation}
    C=D. 
\end{equation}
In addition, the condition
\begin{equation}\label{C1Mink}
    \left.\dot{h}_k\right|_{\eta=0}=0\
\end{equation}
implies
\begin{equation}
    ik(A-B)+2C=0,
\end{equation}
while with
\begin{equation}\label{C2Mink}
   \left.\frac{\partial^3 h_k}{\partial \eta^3} \right|_{\eta=0}=0,
\end{equation}
we obtain: 
\begin{equation}
    ik^3(A-B)+6k^2C=0.
\end{equation}
Combining the two equations, we find  
\begin{equation}
    C=D=0\qquad\text{and}\qquad A=B.
\end{equation}
As a result, our solution for the perturbations becomes: 
\begin{equation}\label{MinkPartSol}
    h_k=A_T\left(e^{ik\eta}+e^{-ik\eta}\right). 
\end{equation}

Therefore, we have obtained only a particular solution for the Minkowski space. Interestingly, comparing the boundary conditions of this case with the dS one, we can notice two crucial differences. First, conditions (\ref{MinkcondNBC}) and (\ref{dSNBCcond}) come with an opposite sign. Second, in the Minkowski case, the third derivative of the metric is vanishing. The reason for both of these results lies in the scale factor. In the dS case, it contributes to the metric coefficients, ultimately changing the overall signature of second-order coefficients, and canceling the third-order coefficient. 

One might wonder if the particular solution could nevertheless be cancelled if one considers higher time derivatives of the metric. Let us consider the n-th time-derivative of the metric perturbation at the boundary: 
\begin{equation}
    \left.\frac{\partial^n h_k}{\partial \eta^n} \right|_{\eta=0}
\end{equation}
We can easily see that if n is odd, and the above expression vanishes in this case,  this will not be enough to remove the remaining constant of integration -- $A_T$. Only possibility that $A_T$ is zero is if even number of derivatives of the metric vanishes at the boundary. We have checked this up to $\mathcal{O}(\eta^2)$ in $H_{\mu\nu}$. In particular, at this order we find: 
\begin{equation}
    g^{(4)}=0+\mathcal{O}\left(\eta^3, \eta^2g_{ij}^{(2)}g^{(2)ij}, \eta^2g^{(2)ij}R_{ij}^{(0)}\right)
\end{equation}
arising from the $H_{00}=0$ condition.From the $H_{ij}=0$ condition we obtain: 
\begin{equation}
    g_{ij}^{(4)}=-D_kD_ig_j^{(2)k}-D_kD_jg_i^{(2)k}+D_kD^kg^{(2)}_{ij}+\mathcal{O}\left(\eta^3, \eta^2g_{ij}^{(2)}g^{(2)ij}, \eta^2g^{(2)ij}R_{ij}^{(0)}\right)
\end{equation}
In terms of the tensor perturbations in the Fourier space, this condition translates to: 
\begin{equation}
    \left.\frac{\partial^4 h_k}{\partial \eta^4} \right|_{\eta=0}=-k^2 \left.\frac{\partial^2 h_k}{\partial \eta^2} \right|_{\eta=0}
\end{equation}
Thus, it is equivalent to the second time derivative of the boundary condition (\ref{MinkcondNBC}), that we have previously obtained. As a result, it will not be able to remove the remaining constant of integration, and with this vanish the tensor modes. 

Nevertheless, the particular solution (\ref{MinkPartSol}) is not natural. It would be far more favourable to have a general one, which would describe the classical plane waves at the linearised level. In the next section, we will investigate how to obtain this general solution.

\subsection{  {\textcolor{Black}{
 \textbf{\textsc{The Alternative Boundary Conditions}}}}}
\normalsize
In the previous section, we found only a particular solution for the tensor modes in the Minkowski case. However, it would be desirable to rather have a general one, where the constants $C$ and $D$ are eliminated.  A hint for this lies in the boundary conditions (\ref{C1Mink}) and (\ref{C2Mink}). We can notice that if we write 
\begin{equation}\label{ABC1}
  \left.-k^2 \dot{h}_k\right|_{\eta=0}=\left.\frac{\partial^3h_k}{\partial \eta^3}\right|_{\eta=0},
\end{equation}
implies $C=0$, so that our solution becomes: 
\begin{equation}
    h_k=A_Te^{ik\eta}+B_Te^{-ik\eta}
\end{equation}
 However, the price we have to pay is the non-vanishing first derivative. In other words, we would need to give up on Maldacena's NBC.

In this subsection, we will show that the \textit{alternative} boundary conditions (ABC) -- (\ref{ABC1}) and (\ref{MinkcondNBC}) -- can be consistently derived from the theory. As before, we will use the following expansion
\begin{equation}
    ds^2=-d\eta^2+\sum_{n=0}^{\infty}\frac{(-1)^n}{n!}\eta^ng^{(n)}_{ij}(\Vec{x})dx^idx^j.
\end{equation}
However, in contrast to the previous analysis, we will require:  
\begin{equation}\label{NoNBC}
    g_{ij}^{(1)}\neq 0, 
\end{equation}
to ensure that the first derivative of metric is non-vanishing at the boundary. The possibility of having (\ref{NoNBC}) with the Starobinsky expansion in the (A)dS case was for the first time introduced in \cite{Grumiller:2013mxa}, in the context of the CG holography. (See also \cite{Irakleidou:2014vla} for further generalization.)

By substituting this expansion into the conditions 
\begin{equation}
    H_{\mu\nu}=0\qquad\text{and}\qquad R=0,
\end{equation}
we find up to the leading order: 
\begin{equation*}
    \begin{split}
    R&=R^{(0)}+g^{(2)}+\frac{1}{4}g^{(1)}g^{(1)}-\frac{3}{4}g^{(1)}_{ij}g^{(1)ij}+\eta\left(g^{(1)}_{ij}R^{(0)ij}+D_iD^{i}g^{(1)}-D_iD_jg^{(1)ij}\right)\\
        &+\eta\left(-g^{(3)}-\frac{1}{2}g^{(2)}g^{(1)}+\frac{5}{2}g^{(1)ij}g^{(2)}_{ij}+\frac{1}{2}g^{(1)ij}g^{(1)}_{ij}g^{(1)}-\frac{3}{2}g^{(1)ij}g^{(1)k}_ig^{(1)_{jk}}\right)\\\\
        H_{00}&=\frac{1}{4}R^{(0)}-\frac{1}{4}g^{(2)}+\frac{1}{16}\left(g^{(1)ij}g^{(1)}_{ij}+g^{(1)}g^{(1)}\right)+\frac{\eta}{4}\left(g^{(1)ij}R^{(0)}_{ij}-D_iD_jg^{(1)ij}+D_iD^ig^{(1)}\right)\\
        &+\frac{\eta}{8}\left(2g^{(3)}-g^{(2)}g^{(1)}-3g^{(1)ij}g^{(2)}_{ij}+g^{(1)ij}g^{(1)k}_ig^{(1)}_{jk}+g^{(1)ij}g^{(1)}_{ij}g^{(1)}\right)
    \end{split}
\end{equation*}
\begin{equation*}
    \begin{split}
            H_{0i}&=\frac{1}{2}\left(D_ig^{(1)}-D_jg^{(1)j}_i\right)\\
        &+\frac{\eta}{2}\left(\frac{3}{2}g^{(1)kl}D_ig^{(1)}_{kl}+\frac{1}{2}g^{(1)}_{ij}D^jg^{(1)}-g^{(1)}_{ij}D_kg^{(1)jk}-g^{(1)jk}D_jg^{(1)}_{ik}+D^jg^{(2)}_{ij}-D_ig^{(2)}\right)\\\\
         H_{ij}&=R_{ij}^{(0)}-\frac{1}{4}g_{ij}^{(0)}R^{(0)}+\frac{1}{2}\left[g^{(2)}_{ij}-g^{(1)}_{jk}g^{(1)k}_i-\frac{1}{2}g^{(0)}_{ij}g^{(2)}+\frac{1}{2}g^{(1)}g^{(1)}_{ij}+\frac{g^{(0)}_{ij}}{8}\left(3g^{(1)}_{kl}g^{(1)kl}-g^{(1)}g^{(1)}\right)\right]\\
    &-\frac{\eta}{4}\left[g^{(0)}_{ij}g^{(1)kl}R^{(0)}_{kl}-g^{(1)}_{ij}R^{(0)}+2D_kD_jg^{(1)k}_i+2D_kD_ig^{(1)k}_j-2D_kD^kg^{(1)}_{ij}-2D_iD_jg^{(1)}\right.\\&\left.+g^{(0)}_{ij}\left(D_kD^kg^{(1)}-D_kD_lg^{(1)kl}\right)\right]+\frac{\eta}{16}\left[-8g^{(3)}_{ij}+4g^{(0)}_{ij}g^{(3)}-4g^{(1)}g^{(2)}_{ij}+8g^{(1)}_{ik}g^{(2)k}_j\right.\\&\left.+8g^{(1)}_{jk}g^{(2)k}_i+2g^{(0)}_{ij}\left(g^{(1)}g^{(2)}-5g^{(1)kl}g^{(2)}_{kl}\right)+g^{(1)}_{ij}\left(g^{(1)}g^{(1)}+g^{(1)kl}g^{(1)}_{kl}\right)-8g^{(1)kl}g^{(1)}_{jk}g^{(1)}_{il}\right.\\&\left.+2g^{(0)}_{ij}\left(3g^{(1)kl}g^{(1)}_{ks}g^{(1)s}_l-g^{(1)kl}g^{(1)}_{kl}g^{(1)}\right)\right]
    \end{split}
\end{equation*}

Let us now study the relations between the metric coefficients arising from (\ref{MinkConditions}), starting with $\mathcal{O}(\eta^0)$. 
 The
 \begin{equation}
     H_{0i}=0
 \end{equation}
 condition implies 
\begin{equation}
    D_ig^{(1)}=D_jg_i^{(1)j}.
\end{equation}
Then, from the conditions
\begin{equation}
    R=0\qquad \text{and} \qquad H_{00}=0,
\end{equation}
we obtain: 
\begin{equation}
    g^{(2)}=\frac{g^{(1)}_{ij}g^{(1)ij}}{2}\qquad\text{and}\qquad R^{(0)}=\frac{1}{4}\left(g^{(1)ij}g^{(1)}_{ij}-g^{(1)}g^{(1)}\right).
\end{equation}
Applying these relations to the
\begin{equation}
    H_{ij}=0,
\end{equation}
we find:
\begin{equation}
    g^{(2)}_{ij}=-2R_{ij}^{(0)}+g^{(1)}_{ik}g^{(1)k}_j-\frac{1}{2}g^{(1)}g^{(1)}_{ij}
\end{equation}
At the linearised level, we can notice that this condition is equivalent to the one that we have obtained with the NBC, (\ref{NBCMink}). The last two terms will give us a contribution that is quadratic in curvature so when we apply this condition to the perturbation theory, we will be able to neglect these terms.

Let us now consider the next order in the conformal time -- $\mathcal{O}(\eta)$. The 
\begin{equation}
    R=0\qquad \text{and} \qquad H_{00}=0,
\end{equation}
conditions imply: 
\begin{equation}
    g^{(3)}=2g^{(1)ij}g^{(2)}_{ij}-g^{(1)ij}g^{(1)}_{jk}g^{(1)k}_i
\end{equation}
and 
\begin{equation}
    g^{(1)ij}R_{ij}^{(0)}=-\frac{1}{2}\left(g^{(1)}g^{(2)}+g^{(1)ij}g^{(2)}_{ij}-g^{(1)ij}g^{(1)}_{jk}g^{(1)k}_i\right).
\end{equation}
Substituting these two relations into 
\begin{equation}
    H_{ij}=0,
\end{equation}
we find: 
\begin{equation}
    \begin{split}
        g_{ij}^{(3)}&=-D_kD_jg^{(1)k}_i-D_kD_ig^{(1)k}_j+D_kD^kg^{(1)}_{ij}+D_iD_jg^{(1)}\\\\
        &+g^{(1)}_{ik}g^{(2)k}_j+g^{(1)}_{jk}g^{(2)k}_i+\frac{1}{2}\left(g^{(1)}_{ij}g^{(2)}-g^{(1)}g^{(2)}_{ij}\right)-g^{(1)kl}g^{(1)}_{jk}g^{(1)}_{il}
    \end{split}
\end{equation}

These relations determine the boundary conditions on the original metric $g_{\mu\nu}$ evaluated at $\eta=0$. Altogether, the relations between the metric at the boundary and the coefficients are given by: 
\begin{equation}
    g_{\mu\nu}^{(0)}=\left.g_{\mu\nu}\right|_{\eta=0}\qquad  g_{\mu\nu}^{(1)}=-\left.g_{\mu\nu}'\right|_{\eta=0}\qquad g_{\mu\nu}^{(2)}=\left.g_{\mu\nu}''\right|_{\eta=0}\qquad g_{\mu\nu}^{(3)}=-\left.g_{\mu\nu}'''\right|_{\eta=0}.
\end{equation}

Let us now apply them to the perturbations of the metric: 
\begin{equation}
    g_{\mu\nu}=\eta_{\mu\nu}+h_{\mu\nu}
\end{equation}
By linearising the above relations in $h_{\mu\nu}$, the previous conditions significantly simplify. In particular, we find: 
\begin{equation}
    \left.\partial_i\dot{h}\right|_{\eta=0}=\left.\partial_j\dot{h}^{j}_i\right|_{\eta=0}\qquad \left.\Ddot{h}^{i}_i\right|_{\eta=0}=0\qquad R^{(0)}=0
\end{equation}
and 
\begin{equation}
   \left. \Ddot{h}_{ij}\right|_{\eta=0}=-2R_{ij}^{(0)}\qquad \left.\frac{\partial^3h_{ij}}{\partial \eta^3}\right|_{\eta=0}=\left(\partial_i\partial_j\dot{h}^k_k+\Delta \dot{h}_{ij}-\partial_k\partial_i\dot{h}_j^{k}-\partial_k\partial_j\dot{h}_i^k\right)_{\eta=0}
\end{equation}
where $\Delta=\partial_k\partial_k$.

Let us now apply these results to the perturbations. \newpage

\textcolor{YellowOrange}{$\diamond$}\;\;\textsc{\textbf{Vector Perturbations}}

 In the synchronous gauge, the gauge-invariant vector perturbations are given by: 
\begin{equation}
    V_i=-\dot{F}_i.
\end{equation}
In this case, the above relations become: 
\begin{equation}
    \left.V_i\right|_{\eta=0}=0\qquad  \left. \dot{V}_i\right|_{\eta=0}=0\qquad   \left.\Ddot{V}_i\right|_{\eta=0}=0
\end{equation}
By applying them to the general solution (\ref{vmodes}), we find that the vector modes vanish in the Minkowski spacetime.

\textcolor{YellowOrange}{$\diamond$}\;\;\textsc{\textbf{Tensor Perturbations}}

In terms of the tensor perturbations 
\begin{equation}
    h_{ij}=h_{ij}^T
\end{equation}
the above conditions become
\begin{equation}
  \left.-k^2 \dot{h}_k\right|_{\eta=0}=\left.\frac{\partial^3h_k}{\partial \eta^3}\right|_{\eta=0},
\end{equation}
and 
\begin{equation}
    \left.\Ddot{h}_k\right|_{\eta=0}=\left.-k^2h_k\right|_{\eta=0}.
\end{equation}
These boundary conditions are precisely the ones that we have intuitively found at the beginning of this section. With them, the general solution for the tensor modes becomes 
\begin{equation}\label{MinkPartSol2}
    h_k=A_Te^{ik\eta}+B_Te^{-ik\eta},
\end{equation}
which is in agreement with the perturbations around the Minkowski space in Einstein's gravity. 

\section{  {\textcolor{Black}{\Large \textbf{\textsc{Summary}}}}}
In this work, we have studied the perturbations of conformal gravity. This theory has six degrees of freedom, out of which two are healthy vector modes, and four tensor ones, among which two are pathological, ghost degrees of freedom.  Due to the conformal invariance, CG cannot distinguish between different conformally flat spacetimes so the presence of ghosts will appear in all of them. Here, we have considered two examples of these spaces -- the de Sitter Universe and Minkowski spacetime.

As it was pointed out in \cite{Maldacena:2011mk}, the ghosts in dS could be removed by a suitable choice of boundary conditions. In particular, the boundary conditions that were chosen have recovered one particular solution of the tensor modes in CG, which was in agreement with perturbations of $E\Lambda$ in the dS space. 

In this paper, we have built on this work by studying the interplay of the boundary conditions that recover general solutions for the perturbations of Einstein's gravity, thus removing the ghost degrees of freedom. 

First, we have considered the de Sitter case. We have found the conditions that reduce the CG to $E\Lambda$ with the GB term. In addition to the previous work \cite{Maldacena:2011mk}, we have shown that these conditions break the conformal invariance and, as a result, select only one of the conformally flat space-times that were allowed by the CG -- the de Sitter Universe. One could wonder if these conditions, imposed on the traceless part of the Ricci tensor and the Ricci scalar could be supplemented with additional requirements, such that the Riemann tensor also takes the appropriate values. We have explored this question in the appendix and found that this is not a possibility. 

Applying the Starobinsky expansion to the conditions, along with the Neumann Boundary Condition, we have derived the boundary condition that should be imposed on the metric. This condition replaces Maldacena's positive frequency mode condition which was selecting only one particular solution for the perturbations. As a result, we find that ghost-like and healthy tensor modes recombine and provide us a fully general solution of the tensor perturbations that agrees with that of the dS space and thus removes the pathological behavior of the conformal gravity.  

Investigating further the case of the Minkowski space-time, we have found that in order to obtain a fully general solution, one has to give up on the Neumann Boundary conditions entirely. In turn, we have derived alternative boundary conditions, with which CG reduces to Einstein's gravity, linearised around the Minkowski space. 

With this approach, we have found that the problematic ghost degrees of freedom can be removed from the conformal gravity, and recover agreement with Einstein's gravity. In future work, we aim to extend this procedure also to other higher-derivative gravity theories and include additional matter sources. 

\begin{comment}
Interestingly, as pointed out in \cite{Maldacena:2011mk}, one would expect that the Minkowski space-time could be obtained from the dS case, by taking the limit $\Lambda\to0$. Here, we see that the limit is seemingly singular. However, we should keep in mind that we have worked with the flat dS coordinates, and thus the limit may be resolved in another frame, such as de Sitter's closed coordinates.

Moreover, in \cite{Chen:2012au} it was pointed out that there is a possibility that the ghosts may re-appear in the presence of external matter. Nevertheless, this may be an unlikely possibility -- the boundary conditions act on the free modes and should thus be taken into account upon the quantization of the theory. 
\end{comment}

\begin{center}
    \large\textsc{\textbf{Acknowledgements}}

\end{center}

\textit{A.H.  would like to thank Misao Sasaki, for enlightening discussions, and for further inspiring the search for general perturbative solutions in the dS space. In addition, A.H. and G.Z. would like to thank Giorgos Anastasiou and George Manolakos for very useful correspondence, 
and the CERN theory department, where part of this work was completed, for hospitality. G.Z. would also like to thank MPP-Munich for hospitality, and MPP-Munich, CERN-TH and DFG Exzellenzcluster 2181:STRUCTURES of Heidelberg University for support.  The work of A.H. is supported in part by the Deutsche Forschungsgemeinschaft (DFG, German Research Foundation) under Germany’s Excellence Strategy – EXC-2111 – 390814868. The work of D.L.~is supported by the Origins Excellence Cluster and by the German-Israel-Project (DIP) on Holography and the Swampland. 
}

\section*{  {\textcolor{Black}{\Large \textbf{\textsc{Appendix}}}}}

\subsection*{\textcolor{YellowOrange}{A.1\;\;}   \textsc{The Riemann tensor and Starobinski Expansion}}
Up to this point, we have considered conditions that specify only the values of the traceless Ricci tensor and the Ricci scalar. Here, we will investigate if setting the Riemann tensor to have an appropriate value could affect these conditions. 

For the de Sitter Universe, the Riemann tensor is given by:
\begin{equation}\label{RiemannEq}
    R_{\mu\nu\rho\sigma}=\frac{\Lambda}{3}\left(g_{\mu\rho}g_{\nu\sigma}-g_{\mu\sigma}g_{\nu\rho}\right)
\end{equation}

By substituting the value of the scale factor for the de Sitter Universe, together with the Starobinsky expansion, we find the following values for the Riemann tensor: 
\begin{equation}
    R_{0000}=0\qquad R_{000i}=0\qquad R_{00ij}=0
\end{equation}
\begin{equation}
    R_{0i0j}=-\frac{1}{H_{\Lambda}^2\eta^4}g_{ij}^{(0)}-\frac{1}{2\eta^2}g_{ij}^{(2)}+\frac{5H_{\Lambda}}{12\eta}g_{ij}^{(3)}-\frac{5H_{\Lambda}^2}{24}g_{ij}^{(4)}+\frac{1}{4}H_{\Lambda}^2g_{ik}^{(2)}g^{(2)k}_j
\end{equation}
\begin{equation}\label{Riemann1}
    R_{0ijk}=\frac{1}{2\eta}\left(D_kg_{ij}^{(2)}-D_jg_{ik}^{(2)}\right)-\frac{H_{\Lambda}}{4}\left(D_kg_{ij}^{(3)}-D_jg_{ik}^{(3)}\right)
\end{equation}
and
\begin{equation}
    \begin{split}
        R_{ijkl}&=\frac{1}{H_{\Lambda}^2\eta^2}R_{ijkl}^{(0)}-\frac{1}{H_{\Lambda}^2\eta^4}(g_{il}^{(0)}g_{jk}^{(0)}-g_{ik}^{(0)}g^{(0)}_{jl})+
        \frac{1}{2}g^{(2)n}_lR^{(0)}_{ijkn}\\&+\frac{1}{4}\left(-D_iD_jg^{(2)}_{kl}-D_iD_kg^{(2)}_{jl}+D_iD_lg^{(2)}_{jk}+D_jD_ig^{(2)}_{kl}-D_jD_kg^{(2)}_{il}+D_iD_jg^{(2)}_{kl}\right)\\
        &-\frac{H_{\Lambda}}{12\eta}\left(-g^{(0)}_{jl}g^{(3)}_{ik}+g^{(0)}_{jk}g^{(3)}_{il}+g^{(0)}_{il}g^{(3)}_{jk}-g^{(0)}_{ik}g^{(3)}_{jl}\right)
    \end{split}
\end{equation}

This should be equated with the r.h.s of (\ref{RiemannEq}). Among all possibilities, only two are non-vanishing.  Expanding them, we find: 
\begin{equation}
    \begin{split}
         H_{\Lambda}^2\left(g_{00}g_{ij}-g_{0i}g_{0j}\right)=-\frac{1}{H_{\Lambda}^2\eta^4}g_{ij}^{(0)}-\frac{1}{2\eta^2}g_{ij}^{(2)}+\frac{H_{\Lambda}}{6\eta}g_{ij}^{(3)}-\frac{H_{\Lambda}^2}{24}g_{ij}^{(4)}
    \end{split}
\end{equation}
and
\begin{equation}
    \begin{split}
        \left(g_{ik}g_{jl}-g_{il}g_{kj}\right)&=\frac{1}{H_{\Lambda}^2\eta^4}(g_{ik}^{(0)}g^{(0)}_{jl}-g_{il}^{(0)}g_{jk}^{(0)})\\&
        +\frac{1}{2\eta}\left(g_{jl}^{(0)}g_{ik}^{(2)}-g_{il}^{(0)}g_{jk}^{(2)}-g_{jk}^{(0)}g_{il}^{(2)}+g_{ki}^{(0)}g_{lj}^{(2)}\right)\\& -\frac{H_{\Lambda}}{6\eta}\left(g_{jl}^{(0)}g_{ik}^{(3)}-g_{il}^{(0)}g_{jk}^{(3)}-g_{jk}^{(0)}g_{il}^{(3)}+g_{ki}^{(0)}g_{lj}^{(3)}\right)
    \end{split}
\end{equation}

Let us now look at the conditions. From (\ref{Riemann1}), we find:
\begin{equation}
    D_kg_{ij}^{(2)}-D_jg^{(2)}_{ik}=0 
\end{equation}

Taking the trace of this relation, we find
\begin{equation}
    D_kg_{ij}^{(2)}=D_jg^{(2)}_{ij}
\end{equation}
which is the relation that comes from the request that $H_{0i}=0$. From the purely spatial part of (\ref{RiemannEq}), we find 
\begin{equation}
    \frac{2}{H_{\Lambda}^2}R_{ijkl}^{(0)}=g_{jl}^{(0)}g_{ik}^{(2)}-g_{il}^{(0)}g_{jk}^{(2)}-g_{jk}^{(0)}g_{il}^{(2)}+g_{ki}^{(0)}g_{lj}^{(2)}
\end{equation}
and 
\begin{equation}
   g_{jl}^{(0)}g_{ik}^{(3)}-g_{il}^{(0)}g_{jk}^{(3)}-g_{jk}^{(0)}g_{il}^{(3)}+g_{ki}^{(0)}g_{lj}^{(3)}=0
\end{equation}

Taking the trace of these equations leads us precisely to the conditions that we have derived before (\ref{CGC1}) and (\ref{CGC2}). However, one component brings us to a problem -- the $ R_{0i0j}$ component leads us to two additional conditions: 
\begin{equation}
    g_{ij}^{(3)}=0\qquad \text{and} \qquad g_{ij}^{(4)}=0
\end{equation}
which set all perturbations, including the tensor ones, to zero. Thus, we can conclude that one cannot set the Riemann tensor to the value analogous to the condition of the Ricci scalar. 

\renewcommand\refname{\textcolor{Black}{\Large\textsc{\textbf{{References\hfill  }}}}}

\end{document}